\documentclass[useAMS,usenatbib]{mn2e}
\usepackage[T1]{fontenc}
\usepackage{prettyref}
\usepackage{amsmath}
\usepackage{graphicx}
\usepackage{subfigure}
\usepackage{color}
\usepackage[breaklinks=true]{hyperref}

\title[Results of two stellar occultations by Ceres]{Results of two multi-chord stellar occultations by dwarf planet (1) Ceres}
\author[A. R. Gomes-J\'unior, B. L. Giacchini, F. Braga-Ribas et al.]{A. R. Gomes-J\'unior$^{1,}$\thanks{E-mail: altair08@astro.ufrj.br},
B. L. Giacchini$^{2,3,4}$,
F. Braga-Ribas$^{5, 6, \dag}$,
M. Assafin$^{1, \dag, \ddag}$,\newauthor
R. Vieira-Martins$^{1,5, \dag, \ddag}$,
J.I.B. Camargo$^{5, \dag}$,
B. Sicardy$^{7}$,
B. Timerson$^{4}$, 
T. George$^{4}$, \newauthor
J. Broughton$^{8}$, 
T. Blank$^{4}$,
G. Benedetti-Rossi$^{5}$, 
J. Brooks$^{4}$,  
R. F. Dantowitz$^{9}$,  \newauthor
D. W. Dunham$^{4}$, 
J. B. Dunham$^{4}$, 
C. K. Ellington$^{4}$,
M. Emilio$^{10}$,
F.R. Herpich$^{11}$,  \newauthor
C. Jacques$^{3, 12}$,
P. D. Maley$^{4,13}$,
L. Mehret$^{10}$,
A.J.T. Mello$^{14}$,
A.C. Milone$^{15}$,\newauthor
E. Pimentel$^{3, 12}$,  
W. Schoenell$^{11}$,
N. S. Weber$^{9}$
\\
$^{1}$Observat\'orio do Valongo/UFRJ, Ladeira Pedro Ant\^onio 43,
CEP 20.080-090 Rio de Janeiro - RJ, Brazil\\
$^{2}$Centro Brasileiro de Pesquisas F\'isicas, Rua Dr. Xavier Sigaud 150, Rio de Janeiro  22290-180, Brazil\\
$^{3}$Se\c{c}\~ao de Oculta\c{c}\~oes/REA-Brasil, Belo Horizonte, Brazil\\
$^{4}$International Occultation Timing Association, P.O. Box 131034, Houston, TX 77219-1034, USA\\
$^{5}$Observat\'orio Nacional/MCTI, R. General Jos\'e Cristino 77, CEP 20921-400 Rio de Janeiro - RJ, Brazil\\
$^{6}$Federal University of Technology - Paran\'a (UTFPR / DAFIS), Rua Sete de Setembro, 3165, CEP 80230-901, Curitiba, PR, Brazil\\
$^{7}$LESIA, Observatoire de Paris, CNRS UMR 8109, Universit\'{e} Pierre et Marie Curie, Universit\'{e} Paris-Diderot, 5 place Jules Janssen,\\ F-92195 Meudon Cedex, France\\
$^{8}$RASNZ Occultation Section, P.O. Box 3181, Wellington, New Zealand\\
$^{9}$Clay Center Observatory at Dexter Southfield, 20 Newton Street, Brookline, MA 02445, USA\\
$^{10}$Universidade Estadual de Ponta Grossa, Ponta Grossa, Brazil\\
$^{11}$Universidade Federal de Santa Catarina, Florian\'opolis, Brazil\\
$^{12}$Centro de Estudos Astron\^omicos de Minas Gerais, Belo Horizonte, Brazil\\
$^{13}$NASA Johnson Space Center Astronomical Society, Houston, Texas, USA\\
$^{14}$Federal University of Technology - Paran\'a (UTFPR / DAELT), Rua Sete de Setembro, 3165, CEP 80230-901, Curitiba, PR, Brazil\\
$^{15}$Instituto Nacional de Pesquisas Espaciais, S\~ao Jos\'e dos Campos, Brazil\\
$^\dag$ Associated to Laborat\'{o}rio Interinstitucional de e-Astronomia - LIneA, Rua Gal. Jos\'e Cristino 77, CEP 20921-400,\\ Rio de Janeiro, Brazil\\
$^\ddag$Affiliated researcher at Observatoire de Paris/IMCCE, 77 Avenue Denfert Rochereau 75014 Paris, France
}
\begin{document}

\date{Submitted: 11 April 2015. Accepted: 11 May 2015.}

\pagerange{\pageref{firstpage}--\pageref{lastpage}} \pubyear{2015}

\maketitle

\label{firstpage}

\begin{abstract}
We report the results of two multi-chord stellar occultations by the dwarf planet (1) Ceres that were observed  from Brazil on 2010 August 17, and from the USA on 2013 October 25. Four positive detections were obtained for the 2010 occultation, and nine for the 2013 occultation. Elliptical models were adjusted to the observed chords to obtain Ceres' size and shape. Two limb fitting solutions were studied for each event. The first one is a nominal solution with an indeterminate polar aspect angle. The second one was constrained by the pole coordinates as given by Drummond et al. Assuming a Maclaurin spheroid, we determine an equatorial diameter of 972 $\pm$ 6 km and an apparent oblateness of $0.08 \pm 0.03$ as our best solution. These results are compared to all available size and shape determinations for Ceres made so far, and shall be confirmed by the NASA's \textit{Dawn} space mission.
\end{abstract}

\begin{keywords}
minor planets, asteroids: individual (1, Ceres) - occultations - planets and satellites: fundamental parameters
\end{keywords}

\section{Introduction}

Ceres is the sole example of a dwarf planet in the inner Solar System. Far from being mere taxonomic information, this suggests the great impact its study can have on the understanding of planetary formation and evolution of the Solar System. Indeed, it was proposed that Ceres' origin could have been as a transneptunian object \citep{McKinnon12}, later scattered to the Main Belt due to the giant planets' migration predicted by the `Nice Model' \citep{Gomes05}. Even if it was formed close to its current location, the dynamical history of the Solar System must have left its signatures on Ceres. These could include not only the late heavy bombardment features that might exist on its surface, but also the makeup of its volatiles, which could have been transported from the outer region.

Since the 1970's it has been speculated that Ceres could contain water vapour, which was recently verified \citep{Kuppers2014}. Although the water regime on this object is still unknown, some internal structure models suggest the existence of a water ice -- or even a liquid water -- layer \citep{CastilloR2011}. Yet the very question of whether Ceres underwent differentiation is open and, on the assumption of an affirmative answer, it is natural to ask if it ever had tectonics, what its geological evolution was, and if it is still active. Inarguably NASA's \textit{Dawn} mission \citep{Russell2004}, which is currently orbiting the dwarf planet, will shed light on several open issues concerning Ceres.

Containing approximately one fifth of the whole Main Belt's mass, Ceres is expected to have an equilibrium figure, i.e. a Maclaurin or a Jacobi ellipsoid. In fact, direct observations of Ceres by means of adaptive optics indicate that it is an oblate spheroid \citep{Drummond2014}. The precise knowledge of its size and shape is of utmost importance, for the models of density, internal structure and differentiation.

The best ground-based technique for determining shape and size of a faraway object is the study of its shadow, cast by a star during an occultation. Since the 1960's occultations have provided measurements of hundreds of asteroids, thanks partially to the fruitful professional-amateur collaboration on the field. More recently, this technique has been applied to objects of the outer Solar System and has unveiled outstanding features of distant bodies, e.g. the ring system around the Centaur (10199) Chariklo \citep{BragaRibas2014}.

The first stellar occultation by Ceres was observed in 1984 \citep{Millis1987} and led to the determination of its size to the precision of some kilometres, at a time when the uncertainties were often ten times larger. The high apparent brightness of Ceres, as compared to most asteroids, imposes a somewhat strong constraint on the stars capable of causing a detectable magnitude drop when occulted. For instance, after the 1984 event, to our knowledge, only four stellar occultations have been observed \citep{Dunham2014}. Two of them had only two chords each, thus not sufficient for providing accurate results\footnote{These events took place on 1994 August 22 and 2010 October 30.}. The two remaining events, which occurred on 2010 August 17 and 2013 October 25, are reported in the present work and are the first ones that used charged-couple devices (CCD) as recording systems. Throughout the paper we shall refer to these events as the `2010 occultation' and the `2013 occultation'.

Both events were predicted by Steve Preston\footnote{Predictions are published at http://asteroidoccultation.com.} on behalf of the International Occultation Timing Association, during routine prediction of asteroidal occultations of bright stars.

This work is organized as follows. In Sections 2 and 3 we analyse the 2010 and the 2013 occultations, respectively. In Section 4 we give the geocentric positions of Ceres derived from the occultations. The comparison of our results to those in the literature is carried out in Section 5.

\section{The 2010 occultation}

\subsection{Observations}

As predicted, on 2010 August 17 Ceres occulted the star TYC 6833-163-1 (UCAC4 313-111823), which has magnitude $V = 11.55$ and has the ICRS position, based on UCAC4 catalogue \citep{Zacharias2013}, to the date of occultation:

\begin{equation}
\left\{ 
  \begin{array}{l l}
    \alpha = 17^{\rmn{h}}18^{\rmn{m}}29\fs0085\\
    \delta = -27\degr 26\arcmin 38\farcs 867
  \end{array}
\right.
\label{Eq: 2010-coord}
\end{equation}

\begin{figure}
\includegraphics[scale=0.42]{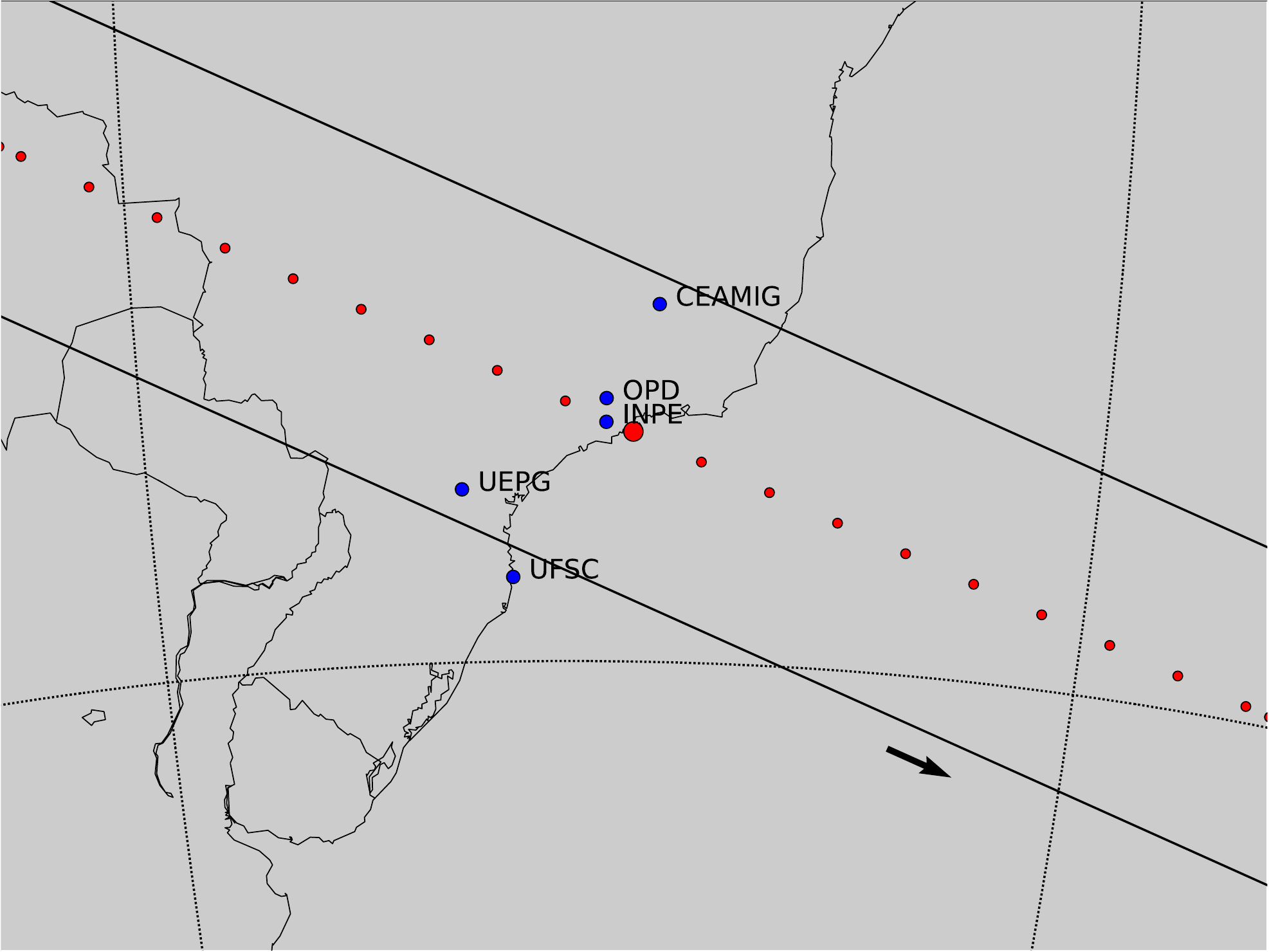} 
\caption{Post-occultation reconstruction of Ceres' shadow path on Earth for the 2010 August 17 event. The big red dot is the geocentric closest approach at 22:40:25 UT. The small red ones represent the centre of the shadow separated by one minute, shadow moves from the left to the right. Blue dots are the sites that have observed the event. As described in text, UFSC had a negative chord.
\label{Fig: Ceres-2010-map}}
\end{figure}

Observations were carried out in Brazil from five different sites as displayed in Table \ref{Tab: obs-2010} and Fig. \ref{Fig: Ceres-2010-map}. The occultation was detected from four of them. The southern-most one (UFSC) had a negative chord. From the positive sites, the one named INPE started observing after the event was already in progress, due to technical difficulties, thus providing only the star's reappearance time; the other three recorded the whole phenomenon.

A remarkable circumstance of this event was the low velocity of Ceres: only 3.9 km s$^{-1}$ in the plane of the sky. Therefore, even exposures of a few seconds would translate in a relevant spatial resolution.

All the observations were made through a sequence of images obtained with CCDs. The times of each exposure were available on the header of each image with an internal accuracy of a few hundredths of a second. CEAMIG had only the integer part of the second available, due to the acquisition software used. The fraction of a second could be retrieved as described on the Section~3.1 of \cite{BragaRibas2013}  (which shall be referred as BR13 hereafter). Cycle times (exposure plus read-out) varied from 2 to 52 seconds, as can be verified on Table~\ref{Tab: obs-2010}, making it a heterogeneous set of observations and imposing an error of a few seconds for the ingress / egress times of some sites.

\begin{table*}
 \centering
 \begin{minipage}{140mm}
  \caption{Circumstances of observation for all observing stations of the 2010 event. \label{Tab: obs-2010}}
  \begin{tabular}{@{}lccccc}
  \hline
          & Longitude & Telescope & Exposure & Result &   \\
     Site & Latitude  & Aperture  & Cycle time & Ingress & Observer\\          
          & Height    & f-ratio & Camera & Egress    & \\          
\hline
 Belo Horizonte & 43$\degr$59\arcmin 51\farcs1~W & LX200 & 5 s & Positive & C. Jacques  \\
 CEAMIG-REA &19$\degr$49\arcmin 49\farcs0~S & 31 cm & 12 s & 22:39:03.9 $\pm$ 0.6 s &  E. Pimentel \\
            & 825 m       &  f/10         &   SBIG ST10        & 22:40:20 $\pm$ 5 s &   \\
 & & & & & \\
 Pico dos Dias    & 45$\degr$34\arcmin45\farcs1~W &  Zeiss     & 1 s & Positive & J. I. B. Camargo \\
 LNA    &22$\degr$32\arcmin03\farcs7~S & 60 cm & 2 s & 22:37:30.3 $\pm$ 0.6 s &  G. B. Rossi \\
            & 1864 m     & f/12.5          &    Andor Ikon       & 22:41:55.3 $\pm$ 0.7 s &              \\
 & & & & & \\
 S\~ao Jos\'e dos       & 45$\degr$51\arcmin44\farcs0~W &  C11  & 2 s & Egress only & A. C. Milone\\
 Campos       &23$\degr$12\arcmin33\farcs0~S & 28 cm & 5 s & Start obs.: 22:39:44 & T. Maldonado\\
 INPE      & 620 m      & f/6.3          & SBIG ST7     & 22:42:03.0 $\pm$ 0.2 s & M. Okada    \\
 & & & & & \\
 Ponta Grossa       & 50$\degr$05\arcmin56\farcs0~W & RCX 400 &30 s & Positive & M. Emilio   \\
 UEPG       &25$\degr$05\arcmin22\farcs2~S & 40 cm & 52 s & 22:37:17 $\pm$ 13 s & L. Mehret   \\
            & 910 m      & f/8          & SBIG STL6E     & 22:39:56 $\pm$ 13 s & \\
 & & & & & \\
 Florian\'opolis       & 48$\degr$31\arcmin20\farcs5~W &   C11      & 3 s & No occultation  & W. Schoenell\\
 UFSC       &27$\degr$36\arcmin12\farcs3~S & 28 cm & 6 s&  Start obs.: 21:49:27 & A. J. T. Mello\\
            & 20 m       & f/10          &  SBIG ST7      &   End obs.: 22:51:21      & F. R. Herpich \\
\hline
\end{tabular}
\end{minipage}
\end{table*}

\subsection{Light curves \label{Sec: light-curve-2010}}

The flux of the star in the five occultation chords was obtained from the FITS images with the \textsc{platform for reduction of astronomical images automatically (praia)} \citep{2011gfun.conf...85A}. The light curves were normalized to the flux of the star plus Ceres, as they were merged right before and after the occultation. Additionally, they were normalized by fitting a polynomial curve (of first or second order) outside the flux drop, so that the flux ratio was set to 1 outside the occultation.

\begin{figure}
\includegraphics[scale=0.58]{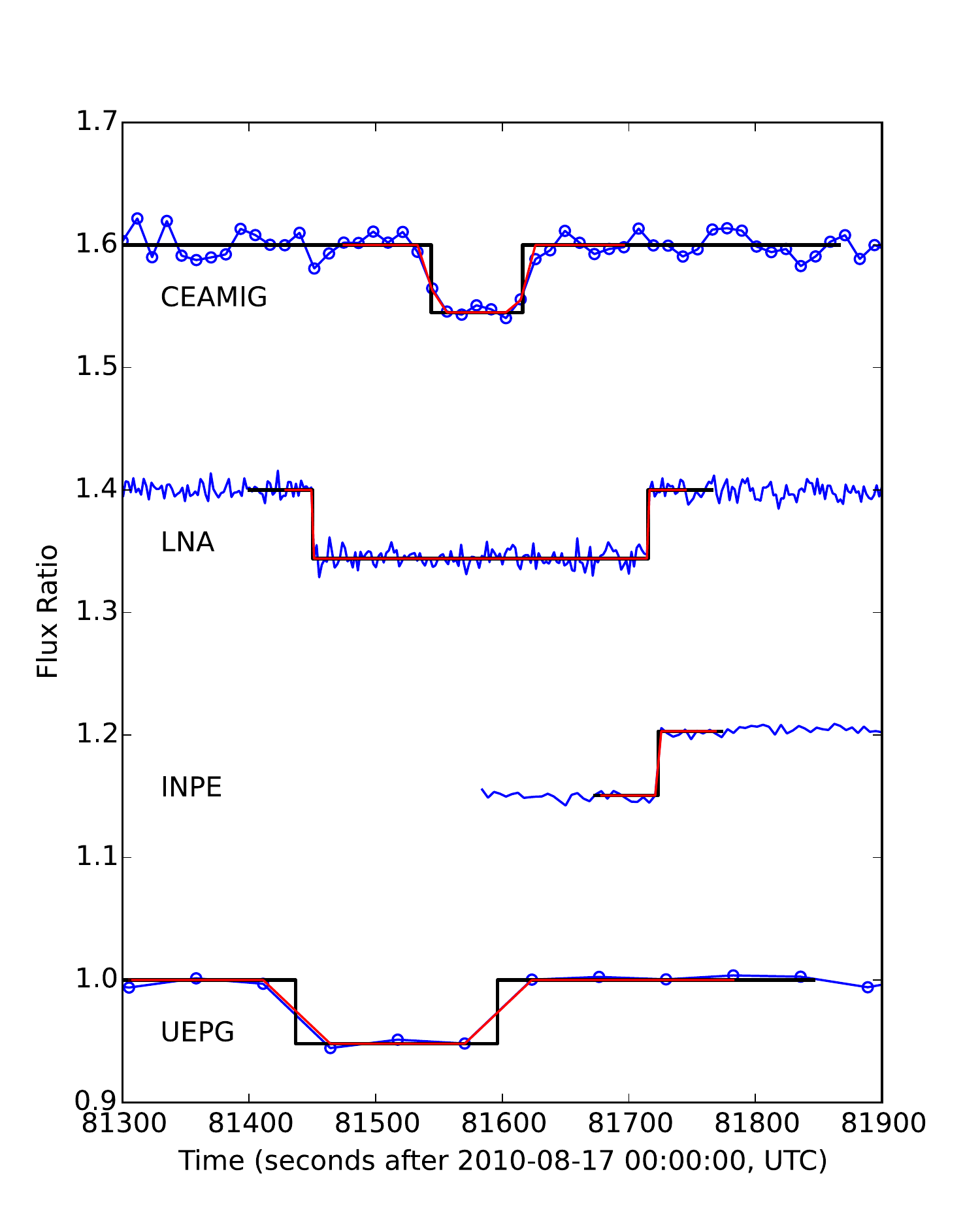} 
\caption{The four occultation light curves normalized and vertically shifted by a factor of 0.2 for better visualisation. The solid black lines are the best fit of the square-well model to the data. Red lines are the square-well model convoluted with the Fresnel diffraction, the star diameter, and the applied exposure time. The mid-times of the occultations do not coincide due to the propagation delays of the shadow due to the distinct longitude of the sites. Exposures at INPE started after the immersion, as explained in the text. \label{Fig: Ceres-2010-curves}}
\end{figure}

The ingress (disappearance) and egress (reappearance) instants of the occultation were obtained for each light curve by fitting a square-well model convoluted with the Fresnel diffraction, the CCD bandwidth, the stellar apparent diameter, and the applied finite exposure time; see \cite{Widemann2009} and BR13. The smallest integration time used in the positive observations was 1.0~s, which translates to almost 3.9 km in the celestial plane. Therefore, the error on the time determination of the ingress and egress is largely dominated by the integration times, not by Fresnel diffraction or star diameter, which are both on the order of a few hundred meters for this event.
The occultation data fit consists in minimizing a classical $\chi^{2}$ function for each light curve, as described in \cite{Sicardy2011} and BR13.\ The free parameter to adjust is the ingress or egress instant, which provides the minimum value of $\chi^{2}$, denoted as $\chi^{2}_{min}$. The best fittings to the 2010 occultation light curves are shown in Fig. \ref{Fig: Ceres-2010-curves}, and the derived occultation times are listed in Table~\ref{Tab: obs-2010}.

\subsection{Limb fitting methodology} \label{Sec: limbfittingmethod}

The methodology used to analyse Ceres' profile from the observations is the same described by \cite{Sicardy2011} and BR13. Each combination of site position and recorded ingress/egress instant, together with star coordinates and Ceres' ephemeris, corresponds to a point in the plane of the sky. The collection of all these points ideally determines the apparent limb of Ceres.

We adopt an elliptic model for the limb profile, resulting from the projection of an oblate spheroid onto the sky plane. This choice is supported by the work of \cite{Drummond2014}, by means of direct imaging of Ceres. Hence, we have $N=7$ chord extremities to adjust the $M=5$ parameters which define an ellipse: apparent semimajor and semiminor axis ($a^\prime$ and $b^\prime$, respectively), position angle $P$ of its semiminor axis and the position $(f_c,g_c)$ of its centre with respect to the occulted star. 
Of course, the apparent semimajor axis {$a^\prime$} is equivalent to the equatorial radius $R_{equa}$ of the ellipsoid.
The coordinates $f_{c}$ and $g_{c}$, in kilometres, are calculated using the JPL\#33 Ceres' ephemeris \citep{Giorgini1996} and the occulted star's position. They are positive toward the local celestial east and north, respectively. The position angle $P$ is counted positively from the direction of local celestial north to celestial east. The apparent oblateness can be defined by $\epsilon^\prime = 1 - (b^\prime/a^\prime)$. The best-fitting solution is obtained minimizing a reduced $\chi^{2}_{r}$ function, where we define the number of degrees of freedom of the problem as $\mathcal{N} \equiv N - M$.  All the procedures that allow the determination of the error bars of the physical parameters can be found on BR13.

\subsection{Limb fitting solutions}\label{Sec: limbfitting-2010}

Two possible solutions were considered for the limb fitting. The first, which we call nominal solution, consists of determining the five parameters that characterize an ellipse from the seven observed contacts. As we shortly show, it led to a rather large uncertainty on the position angle. Furthermore, the nominal solution alone is not capable of returning the true oblateness, which can be evaluated through equation 2 of BR13  provided that the aspect angle $\zeta$ is known.

It is possible to derive the pole position $(\zeta, P)$ from the coordinates of Ceres' pole $(\alpha_{p}, \delta_{p})$ and its ephemeris $(\alpha, \delta)$ via
\begin{equation}
\left\{
\begin{array}{l l}
\cos(\zeta)= -\sin(\delta_{p})\sin(\delta) - \cos(\delta_{p})\cos(\delta)\cos(\alpha_{p}-\alpha)\\
\tan(P) = \dfrac{\sin(\alpha_{p}-\alpha)} {\tan(\delta_{p})\cos(\delta) - \sin(\delta)\cos(\alpha_{p}-\alpha)},\\
\end{array}
\right.
\label{Eq: pole-position}
\end{equation}
where $\zeta = 0\degr$ and $\zeta=90\degr$ respectively correspond to a pole-on and an equator-on geometry.

From observations with adaptive optics spanning a 10-year period, \cite{Drummond2014} determined the position of Ceres' polar axis within an error of only 3$\degr$:
\begin{equation}
\alpha_{p} = (287 \pm 3) \degr, \quad 
\delta_{p} = (+64 \pm 3) \degr,
\label{Eq: Pole}
\end{equation}
in equatorial J2000 coordinates. Together with Ceres' ephemeris at the moment of the occultation, this corresponds to the polar aspect angle $\zeta = 86.1\degr$, which is very close to an equator-on geometry. Hence, we expect true figures to be similar to apparent ones.

The knowledge of Ceres' pole not only allows the determination of its polar aspect, it suffices to set its position angle. Therefore \textbf{Eqs. \ref{Eq: pole-position} and \ref{Eq: Pole}} may act as a constraint for $P$, and a second solution can be obtained by probing the parameter space with the restriction that the position angle is confined to the range that follows from Eq. \ref{Eq: Pole}. We call this the `pole-constrained solution'.

\subsubsection{Nominal solution}

With the seven observed contacts it is possible to adjust the five parameters which define an ellipse. For the best-fitting solution we find $\chi^2_{r,min} = 0.24$, which could be interpreted as a slightly overestimation of the error bars with regard to the good quality of the fit. However, inasmuch as the problem has only two degrees of freedom, it is far from the statistical realm and relatively small $\chi^2_{r,min}$ are acceptable.

The resulting values of equatorial diameter, oblateness, position angle and centre coordinates are presented in the second column of Table~\ref{Tab: results}.

Already mentioned, the parameter with the largest uncertainty is the position angle: spanning on a 20$\degr$ interval, its determination has a relative precision worse than 10 per cent. Clearly, the coordinates of Ceres' pole (Eq. \ref{Eq: Pole}) can impose a strong constraint on the position angle, as the next solution shows.

Finally, the correction to the oblateness due to Ceres' polar aspect angle lies within the 1-sigma error bar and has no statistical relevance; hence $\epsilon = 0.08 \pm 0.03$.

\subsubsection{Pole-constrained solution}

\begin{figure}
\includegraphics[scale=0.37, angle=-90]{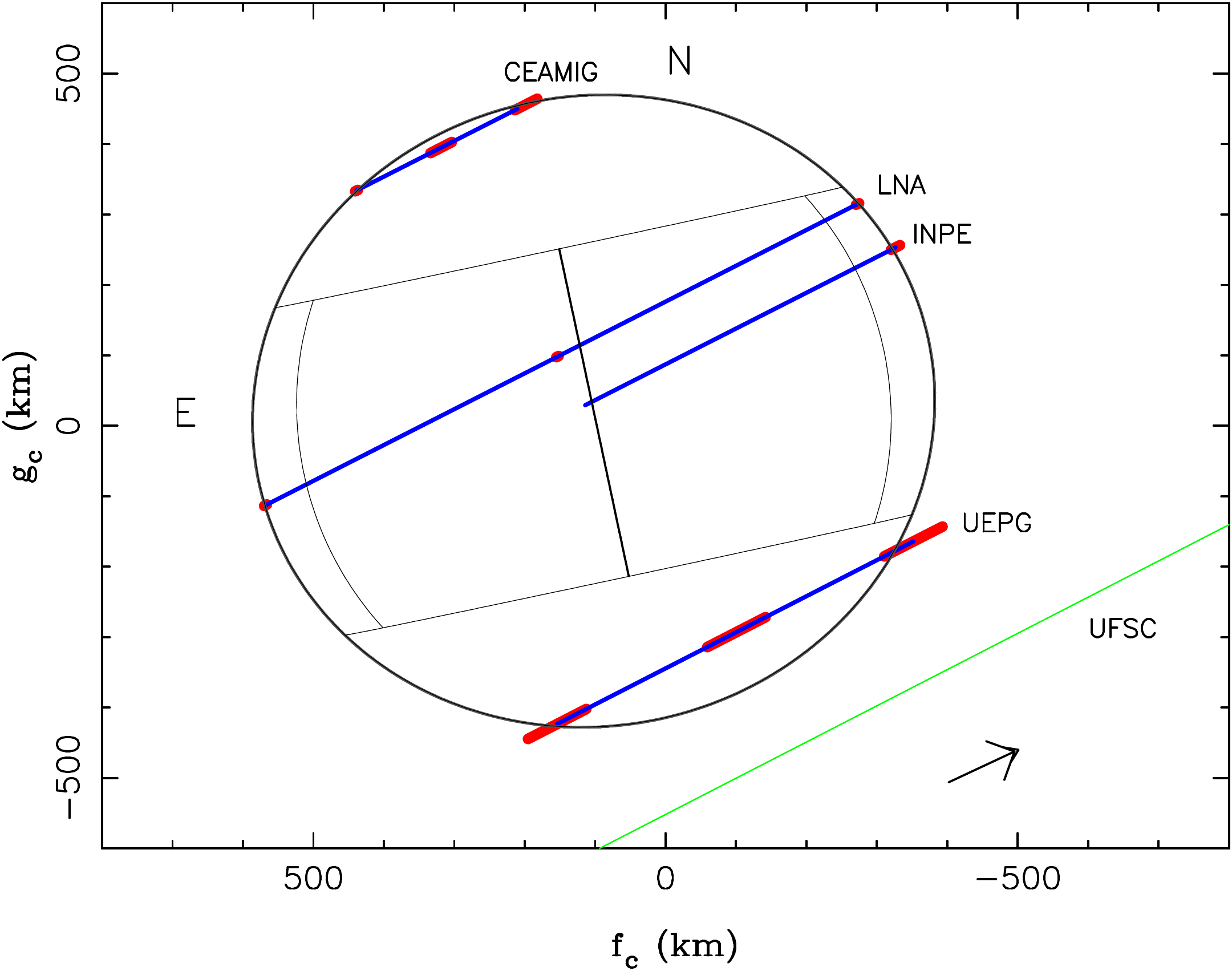} 
\caption{The best elliptical fit for the occultation chords for the event of 2010 using the times from Table~\ref{Tab: obs-2010} and the pole-constrained solution. The arrow indicates the direction of motion, blue lines are the observed chords, the red segments are the ingress, egress and mid-occultation error bars at 1-$\sigma$ level. \label{Fig:Ceres-2010-body}}
\end{figure}

At the occultation, the coordinates (Eq. \ref{Eq: Pole}) of Ceres' rotational pole correspond to the position angle $P = (12 \pm 3)\degr$. Exploration of the parameter space, restricted to ellipses with position angles laying within this range, results in the pole-constrained solution. The related physical parameters are displayed on the third column of Table \ref{Tab: results} in boldface, while the best-fitting solution is depicted in Fig.\ref{Fig:Ceres-2010-body}.

We notice that the constraint corresponds to the upper limit of nominal solution's 1-sigma error bar for $P$. On the other hand, it selects the smallest values of semimajor axis, improving its determination by a factor of about 2. Notwithstanding, oblateness' figures remain the same.

\begin{table*}
 \centering
 \begin{minipage}{140mm}
  \caption{Results of limb fitting to the data of the 2010 and 2013 events.\label{Tab: results}}
  \begin{tabular}{@{}lcccc}
  \hline
     Solution & 2010/Nominal & \textbf{2010/Pole-constrained} & 2013/Nominal & 2013/Pole-constrained \\
\hline
Equatorial diameter (km) & 982 $\pm$ 14 & \textbf{972 $\pm$ 6}  & 971 $\pm$ 7  & 971 $\pm$ 7\\
True oblateness        & 0.08 $\pm$ 0.03 & \textbf{0.08 $\pm$ 0.03} & 0.08 $\pm$ 0.04 & 0.08 $\pm$ 0.04\\
Position angle (deg)   & 5 $\pm$ 10    & \textbf{12 $\pm$ 3} (*)& 22 $\pm$ 5    & 25 $\pm$ 3 (*)\\
$f_c$ (km)             & 97 $\pm$ 9   & \textbf{102 $\pm$ 5}   & 77 $\pm$ 6    & 78 $\pm$ 6\\
$g_c$ (km)             & 16 $\pm$ 15  & \textbf{21 $\pm$ 11}  & 13 $\pm$ 16   & 13 $\pm$ 16\\
$\chi^2_{r,min}$       & 0.24          &  \textbf{0.42}         & 1.27          & 1.27\\
\hline
\end{tabular}
\textbf{Notes.} In bold we highlight our best solution. Error bars are at 1-sigma level. The polar diameter ($D_{pol}$) can be easily calculated from $D_{pol}=D_{equa}(1 - \epsilon)$.
(*) Position angle derived from Ceres' rotational pole coordinates determined by \cite{Drummond2014}.
\end{minipage}
\end{table*}

\section{The 2013 occultation}

\subsection{Observations}\label{Sec: observation-2013}

\begin{figure}
{\includegraphics[scale=0.42]{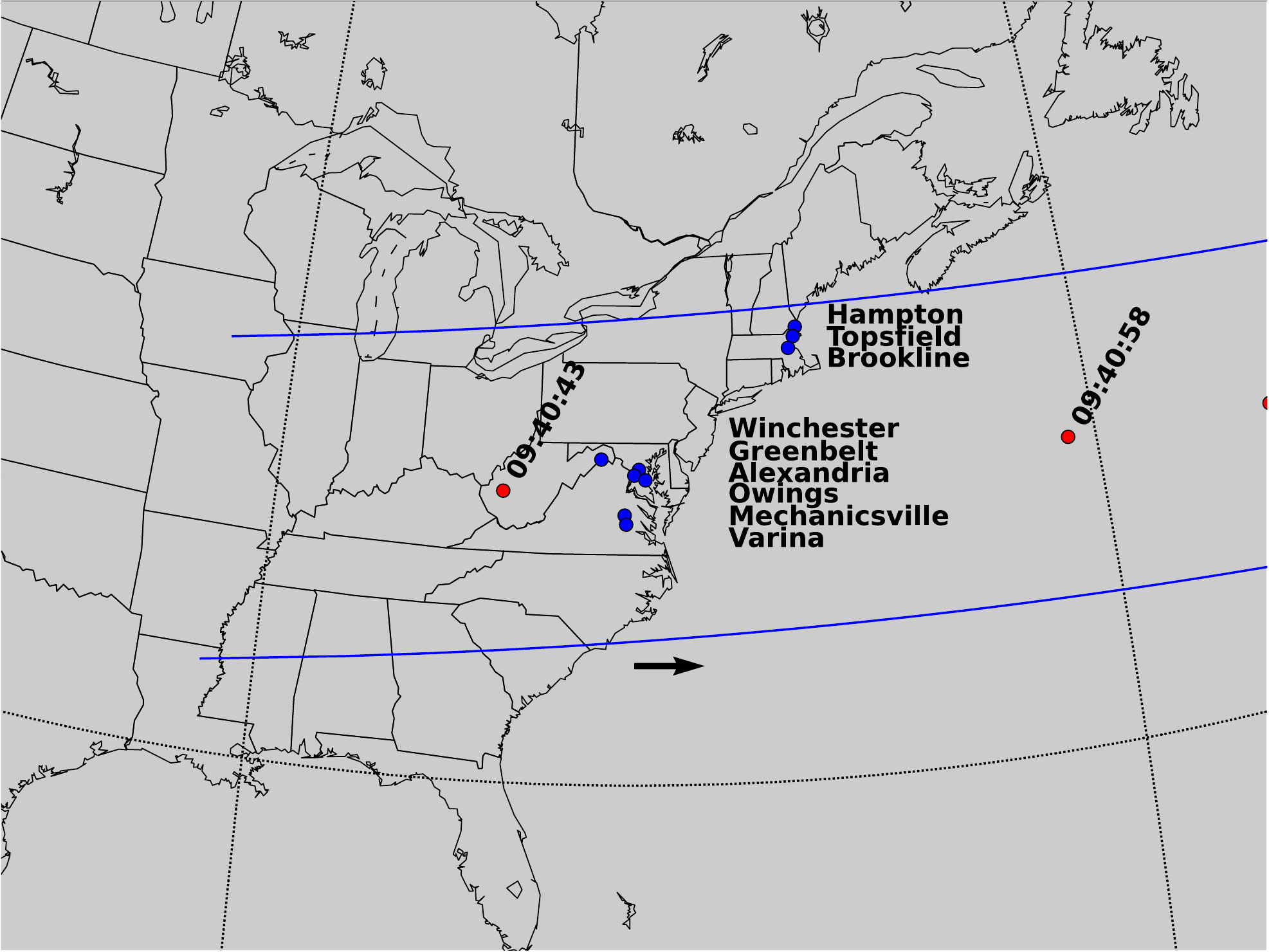}}
\caption{Post-occultation reconstruction of Ceres' shadow path on Earth for the 2013 October 25 event at the east coast of USA. 
Upper view of the occultation over the sites that observed the event (blue dots). Red points are the centre of the shadow separated by 15 seconds.
\label{Fig: Ceres-2013-map}}
\end{figure}

The event which took place on 2013 October 25 involved the star TYC 865-911-1 (UCAC4 496-058191), of magnitude $V = 10.05$. Based on UCAC4 \citep{Zacharias2013}, the ICRS position to the date of occultation is:
\begin{equation}
\left\{ 
  \begin{array}{l l}
    \alpha = 11^{\rmn{h}}57^{\rmn{m}}52\fs7641\\
    \delta = +09\degr 07\arcmin 49\farcs835
  \end{array}
\right.
\end{equation}
The occultation could only be visible in the United States, before dawn, as depicted in Fig. \ref{Fig: Ceres-2013-map}.

Nine positive chords were obtained by the variety of instruments listed in Table \ref{Tab: obs-2013}. Each station was equipped with a video camera with negligible readout times. This was of particular importance, since in this event Ceres' shadow speed was 42.6 km s$^{-1}$, much faster than in the 2010 event.

Three different timing synchronization procedures were adopted among the set of observing stations. At Greenbelt and Owings the 1PPS signal of a GPS unit was used to calibrate time stamps which were inserted at each frame of the video. Time extraction is thus straightforward, after taking camera delays into account. On the other hand, at Brookline the clock would be synchronized by an internet server. A lack of connection, however, resulted in spurious times. In fact, comparison between the times obtained at this station and the others suggests that the former have a delay of about 64 s. Therefore, we do not use Brookline's 
times in the analysis that follows. Finally, at the remaining six stations the videos were recorded by camcorders on digital tapes. The timing method consisted in the comparison of the camcorder internal clock to a 1PPS GPS signal, before and after the recording of the occultation. Absolute timing errors of this procedure are expected to be less than 0.1 s.

During the event Ceres was low in the horizon, with altitudes between $15\degr$ (Winchester) and $20\degr$ (Hampton). Strong scintillation is expected in such a scenario which, combined to short integration times and the low magnitude drop of the event, resulted in rather noisy light curves and thus larger uncertainties in the time of the contacts, as is shown in the next section.

\begin{table*}
 \centering
 \begin{minipage}{140mm}
  \caption{Circumstances of observation for the observing stations of the 2013 event.\label{Tab: obs-2013}}
  \begin{tabular}{@{}lccccc}
  \hline
          & Longitude & Telescope: & Camera  & Result &   \\
     Site & Latitude  & Aperture  & Cycle time & Ingress & Observer\\          
          & Height    & f-ratio    &           & Egress    & \\          
\hline
 Hampton & 70$\degr$48\arcmin59\farcs7~W & 12 cm & PC164C-EX2 & Positive & T. Blank \\
  &42$\degr$53\arcmin52\farcs8~N & f/5 & 0.033 s     & 09:40:46.9 $\pm$ 0.1 s &   \\
            & 7 m       &  &     & 09:40:57.26  $\pm$ 0.08 s &   \\
 & & & & & \\
 Topsfield & 70$\degr$55\arcmin16\farcs6~W & 12 cm & PC164C-EX2 & Positive & T. Blank \\
  &42$\degr$37\arcmin55\farcs9~N & f/5      & 0.033 s   & 09:40:45.4 $\pm$ 0.1 s &   \\
            & 45 m      &  &     & 09:40:58.0  $\pm$ 0.1 s &   \\
 & & & & & \\
 Brookline & 71$\degr$08\arcmin14\farcs5~W & 64 cm & Infinity2-1R & Positive & N. Weber \\
  &42$\degr$18\arcmin27\farcs4~N & f/9.6 & 0.015 s     & 09:41:48.00 $\pm$ 0.01 s &  R. Dantowitz  \\
            & 109 m     &       &     & 09:42:02.93 $\pm$ 0.01 s   \\
 & & & & & \\
 Winchester & 78$\degr$14\arcmin39\farcs6~W & 36 cm & PC164C & Positive & J. Brooks \\
  &39$\degr$16\arcmin21\farcs5~N & f/5 & 0.033 s    & 09:40:33.26 $\pm$ 0.08 s &   \\
            & 211 m     &  &     & 09:40:55.86  $\pm$ 0.09 s &   \\
 & & & & & \\
 Greenbelt & 76$\degr$52\arcmin09\farcs4~W & 12 cm & PC164C-EX2 & Positive & J. Dunham \\
  &38$\degr$59\arcmin12\farcs1~N & f/2.5 & 0.033 s    & 09:40:33.6 $\pm$ 0.1 s & D. Dunham  \\
            & 52 m      &   &     & 09:40:56.4  $\pm$ 0.1 s &   \\
 & & & & & \\
 Alexandria & 77$\degr$02\arcmin28\farcs3~W & 7 cm & Watec120N & Positive & P. Maley \\
  &38$\degr$49\arcmin19\farcs1~N & f/10 &  0.067 s   & 09:40:33.3 $\pm$ 0.1 s &   \\
            & 8 m       &   &     & 09:40:56.1  $\pm$ 0.1 s &   \\
 & & & & & \\
 Owings & 76$\degr$38\arcmin06\farcs3~W & 25 cm & PC164C  & Positive & C. Ellington \\
  &38$\degr$41\arcmin26\farcs5~N & f/3.3 &  0.033 s & 09:40:34.27 $\pm$ 0.05 s &   \\
            & 38 m                &       &     & 09:40:56.0  $\pm$ 0.2 s &   \\
 & & & & & \\
 Mechanicsville & 77$\degr$23\arcmin06\farcs7~W & 12 cm & PC164C-EX2 & Positive & D. Dunham \\
  &37$\degr$41\arcmin26\farcs1~N & f/2.5 & 0.033 s    & 09:40:33.0 $\pm$ 0.1 s &   \\
            & 60 m      &   &     & 09:40:54.8  $\pm$ 0.1 s &   \\
 & & & & & \\
 Varina & 77$\degr$19\arcmin49\farcs3~W & 12 cm & PC164C-EX2 & Positive & D. Dunham \\
  &37$\degr$25\arcmin58\farcs6~N & f/2.5 &  0.033 s   & 09:40:32.4 $\pm$ 0.3 s &   \\
            & 19 m      &   &     & 09:40:53.1  $\pm$ 0.2 s &   \\

\hline
\end{tabular}
\end{minipage}
\end{table*}

\subsection{Light curves}

All videos were converted to FITS images and the photometry of the target was obtained via \textsc{praia} \citep{2011gfun.conf...85A}. The light curves were normalized to the flux of a reference star when available on the field.

To reduce the noise, the data were binned by groups of five images -- with the exception of  Greenbelt, which was averaged in sets of ten. This procedure enlarges the effective integration times reported in Table~\ref{Tab: obs-2013} by a factor of 5 (or 10). As in the 2010 event, an additional normalization by a polynomial curve was applied.

The ingress and egress instants of the occultation were obtained by the same procedure explained in Section \ref{Sec: light-curve-2010}. 
Since the typical effective integration time used (0.17~s) translates to about 7 km in the celestial plane, and the Fresnel scale and star diameter are again of the order of a few hundreds of meters, the theoretical occultation light curves are largely dominated by the integration times, as for the 2010 event.

\begin{figure}
\includegraphics[scale=0.58]{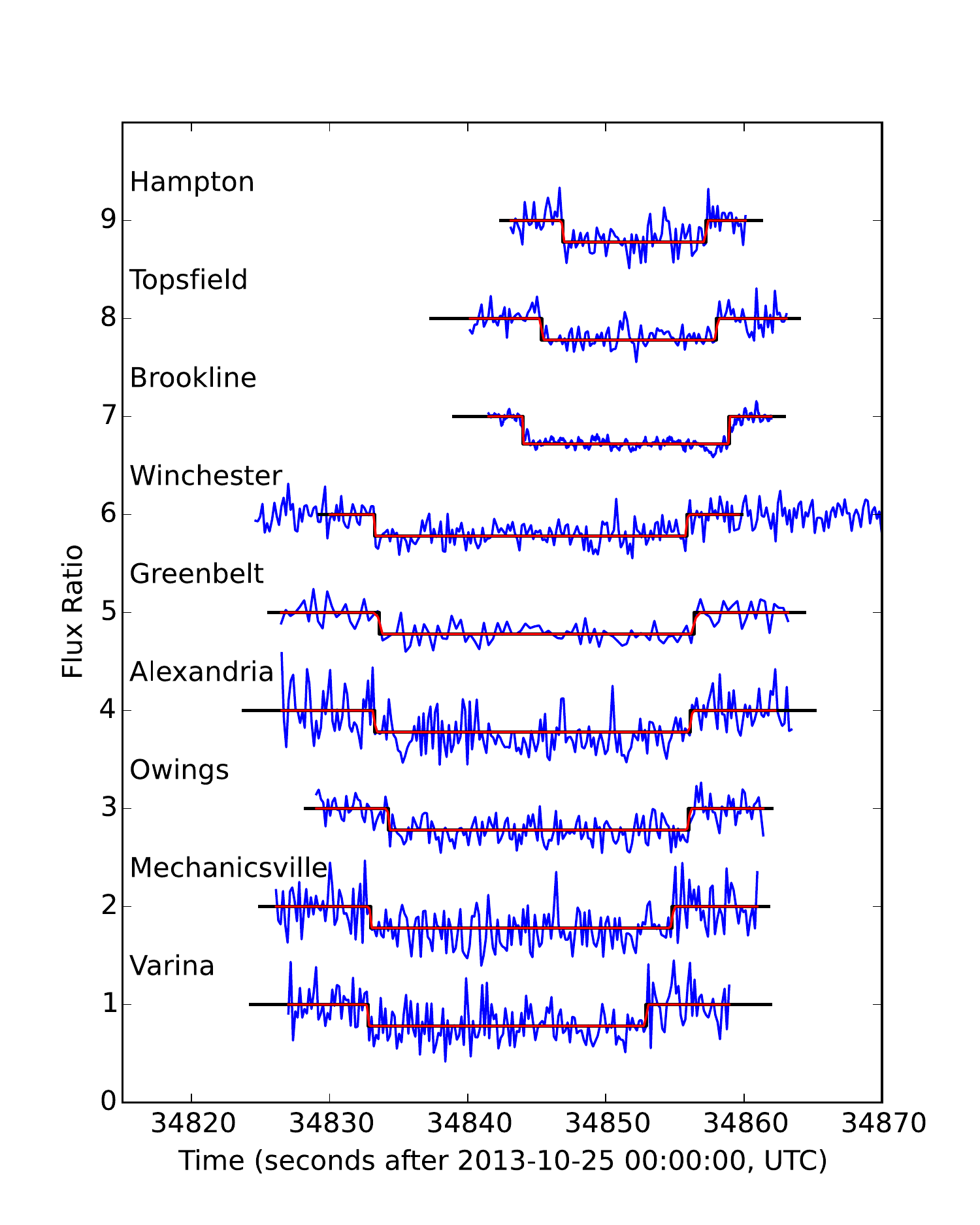} 
\caption{The nine occultation light curves normalized and vertically shifted by a factor of 1.0 for better viewing, see Fig.~\ref{Fig: Ceres-2010-curves} for the explanation of the graph. The light curve of Brookline is shifted by -64 s as explained in the text. \label{Fig: Ceres-2013-curves}}
\end{figure}

The best-fittings to the occultation light curves are shown in Fig.~\ref{Fig: Ceres-2013-curves}, and the derived occultation times are listed in Table~\ref{Tab: obs-2013}.

\subsection{Limb fitting solutions}

\begin{figure}
\includegraphics[scale=0.36]{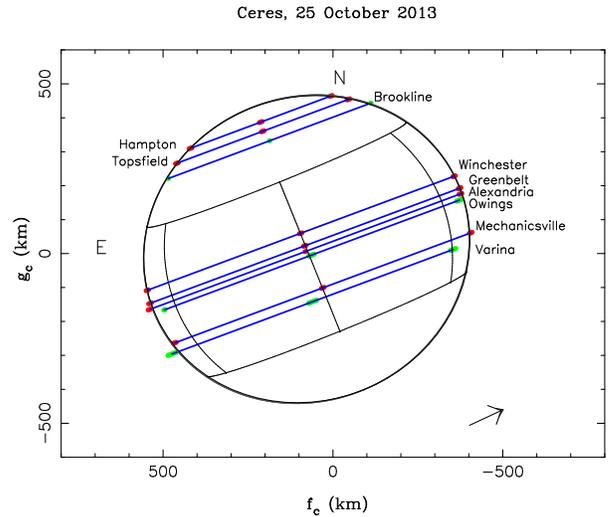}
\caption{The best elliptical fit for the occultation chords for the event of 2013 using timing from Table \ref{Tab: obs-2013} and the pole-constrained solution.  The arrow indicates the direction of motion, blue lines are the observed chords, the red and green segments are the ingress, egress and mid-occultation error bars at 1-$\sigma$ level. The chords with green error bars were not used during the limb fit process. The chord of Brookline if shifted by -64s as explained in the text.\label{Fig: Ceres-2013-body}}
\end{figure}

Elliptic limb profiles were adjusted to all the available\footnote{Brookline's chord was not used for limb fitting since it had an inaccurate absolute time.} chords by the same procedure described in Section \ref{Sec: limbfittingmethod}. This yielded $\chi^2_{r,min} = 13$, suggesting that an elliptic model is not satisfactory to the data. Indeed, a quick glance at a plot of the observed chords, such as in Fig. \ref{Fig: Ceres-2013-body}, shows that the Varina chord seems to be somewhat advanced with respect to the others. Taking into account that in this station time stamps were not inserted on the video frames, it is fairly possible to attribute this advance to an eventual problem on the correspondence between camcorder's and GPS' times. This could be caused, for example, if the camcorder delayed to start the recording and, since this was an unattended pre-pointed station, this fact would not be noticed by the analysis of the video itself.

The immersion recorded at Owings also seems to be shifted (delayed) with respect to the nearby chords (see Fig. \ref{Fig: Ceres-2013-body}). This chord, actually, has roughly the same length of Mechanicsville's, despite the fact that they are separated by about 100 km. Differently from Varina, though, this station had time stamps inserted in each frame of the video, which makes it more unlikely to justify an eventual timing issue. Another possible explanation for the apparent problem of the times of this chord is the determination of the ingress and egress instants in the light curve analysis, which could have been affected by noise. Finally, the delay during the ingress could be caused by a relief feature in Ceres; we shall soon return to this hypothesis.

In a second limb fitting, thus, we did not consider Brookline, Varina and Owings chords. The adjust of the five parameters which define an ellipse to the 12 contacts then resulted in $\chi^2_{r,min} = 1.27$, indicating that the fitting is in good agreement with the observed data within the error bars. This is the solution depicted in Fig.~\ref{Fig: Ceres-2013-body}, where we see that the chord length measured in Brookline is compatible to the model. The associated physical parameters are presented in Table~\ref{Tab: results} as the nominal solution.

For this event the polar aspect angle is $\zeta = 90.7\degr$, which makes the true oblateness equal to the apparent one, within the error bars.

The position angle of this nominal solution is better constrained than the 2010 occultation. Actually, the uncertainty in the former is of $5\degr$ in contrast to $10\degr$ of the latter. This suggests that the pole-constrained solution obtained via the position of Ceres' pole (Eq. \ref{Eq: Pole}) would not be significantly different from the nominal one.

This assumption was confirmed when we carried out the limb fitting with the constraint $P = (25 \pm 3)\degr$, the position angle at the occultation which follows from Eq. \ref{Eq: Pole}. The physical parameters related to this pole-constrained solution are presented in the last column of Table~\ref{Tab: results}, and are essentially the same of the nominal solution.

In the hypothesis that the delay observed in the immersion at Owings could be associated to a limb topography feature, the recorded contact would then correspond to a negative elevation of 31 $\pm$ 4 km with respect to the best-fitting ellipse. Theoretical models, however, predict that reliefs in Ceres should not be higher than about 10--20 km \citep{Johnson1973}, while published observational data sets the bound of 18 km \citep{Carry2008}. More recently, images by the probe \textit{Dawn} also reveal an even smoother surface. Therefore, the association of Owings first contact to a relief is improbable.

\section{Astrometry from Occultations}

The objects' geocentric positions derived from stellar occultations are most valuable for improving their orbits \citep{Desmars2015}. Usually the precision of the astrometric positions are limited by the accuracy of the occulted star position, not by the limb fit. Ceres' geocentric J2000 positions at the time of each occultation are displayed in Eqs.~\ref{Eq: Ceres-2010_pos} and \ref{Eq: Ceres-2013_pos}. The errors of the positions come from the errors of the star positions, taken from the catalogues and from the errors of the relative apparent distances between star and Ceres, derived from the limb fit (which are displayed as $\Delta\alpha \cos\delta$, $\Delta\delta$).

\begin{equation}
2010 ~ \text{Aug} ~ 17
\left\{
 \begin{array}{l l}
    \text{Time} = 22:40:00\\
    \alpha = 17^{\rmn{h}}18^{\rmn{m}}29\fs0122\pm0\farcs027\\
    \delta = -27\degr 26\arcmin 38\farcs617\pm0\farcs028\\
    \Delta\alpha\cos\delta = 0\farcs003; ~~\Delta\delta = 0\farcs007
  \end{array}
\right.
\label{Eq: Ceres-2010_pos}
\end{equation}
\begin{equation}
2013 ~ \text{Oct} ~ 25
\left\{
  \begin{array}{l l}
    \text{Time} = 09:45:00\\
    \alpha = 11^{\rmn{h}}57^{\rmn{m}}52\fs9154\pm0\farcs019\\
    \delta = +09\degr 07\arcmin 49\farcs865\pm0\farcs021\\
    \Delta\alpha\cos\delta = 0\farcs002; ~~\Delta\delta = 0\farcs007
  \end{array}
\right.
\label{Eq: Ceres-2013_pos}
\end{equation}

\section{Discussion}

A quick glance at Table~\ref{Tab: results} shows an overall agreement between the physical parameters derived from both occultations, specially in the equatorial diameter. The differences of the solutions occur basically on the size of their error bars, and can be justified by the particularities of each set of data, as discussed below. 

The 2010 event, for example, had only seven contacts; none the less, they were well distributed over Ceres' disc (see Fig. \ref{Fig:Ceres-2010-body}) acting as a constraint to its shape. On the other hand, the 2013 event had five more exploitable contacts, but they were concentrated in certain regions of the body. In particular, the absence of chords close to Ceres' south pole made its oblateness less well determined here than in the 2010 event.

However, even our best measurement for the oblateness, $\epsilon=0.08 \pm 0.03$, has still a higher uncertainty with regard to other figures published in the literature, as Table~\ref{Tab: Ceres-final} shows. A larger number of uniformly spaced chords would be necessary to offer a best constraint to the oblateness. 

The few chords of the 2010 occultation could themselves only constrain the position angle of the object to a uncertainty of $10\degr$. This uncertainty was reduced by a factor of two in the 2013 event, approaching --and verifying-- the result predicted by the work of \cite{Drummond2014}. As was shown, using the coordinates of Ceres' polar axis to limit the position angle was not an efficient procedure in the 2013 event, in the sense that it did not result in significant changes in the parameters obtained in the nominal solution (see Table \ref{Tab: results}).

On the other hand, constraining the position angle on the 2010 occultation was proved to reduce the error bars of the other parameters (disregarding oblateness). Moreover, this procedure resulted in an excellent agreement between the equatorial radius figures of both events.

The 2013 occultation, therefore, offers not only an independent verification of the figures resulted from the 2010 event, but also validates the procedure carried out there which led to the best-constrained parameters in this work.

Comparison of Ceres' equatorial diameter as measured by different techniques is carried out in Table \ref{Tab: Ceres-final}. We note an overall agreement between our result to those obtained via direct imaging by the Hubble Space Telescope (HST) \citep{Thomas2005}, the Keck Observatory and the ESO VLT \citep{Drummond2014}. The smaller figure reported by \cite{Carry2008} may be justified by the fact that in this study the effects of limb-darking were not taken into account, as pointed out by \cite{Drummond2014}.

As mentioned in the Introduction, the 1984 event \citep{Millis1987} is the only occultation data to which we can compare our result. The measured diameters do not agree, ours being larger by 2-sigma. It is difficult to state for sure the reasons for this divergence. One way to clarify the issue is to redetermine the immersion and emersion instants from the original light curves using the same methodology presented in our work. Note that in this occultation a variation of 0.5 s in the contact times correspond to almost 7 km on Ceres' limb, which is on the same order of the residuals of their elliptical fit. However, we could not do that, as we had no access to the original 1984 light curve data. Moreover, we have no description of how the immersion and emersion instants of the chords were actually determined from these light curves in \cite{Millis1987}.

NASA's \textit{Dawn} mission shall bring light to these questions, which will be important not only for the knowledge of Ceres itself, but also for all the techniques used so far to study the physical properties of small Solar System objects, such as the stellar occultations.

\begin{table}
  \caption{Ceres' equatorial diameter and oblateness \label{Tab: Ceres-final}}
  \begin{centering}
  \begin{tabular}{@{}lccc}
  \hline
     Eq. diameter (km) & Oblateness & Method & Ref. \\
\hline
$972 \pm 6$  & $0.08  \pm 0.03$  & Occultation & 1\\
$967 \pm 10$ & $0.078 \pm 0.015$ & Keck+VTL    & 2\\
$959 \pm 5$  & $0.074 \pm 0.007$ & Keck        & 3\\
$975 \pm 4$  & $0.067 \pm 0.005$ & HST         & 4\\
$959 \pm 5$  & $0.05  \pm 0.01$  & Occultation & 5\\
\hline
\end{tabular}
\par\end{centering}
\textbf{References.} 1: Present work. 2: \cite{Drummond2014}. 3: \cite{Carry2008}. 4: \cite{Thomas2005}. 5: \cite{Millis1987}.
\end{table}

\section*{Acknowledgements}

ARGJ thanks the financial support of CAPES. BLG thanks CNPq. F.B.R. acknowledges PAPDRJ-FAPERJ/CAPES E-43/2013 number 144997, E-26/101.375/2014 and CDFB-CAPES/Brazil. M.A. acknowledges CNPq grants 473002/2013-2, 482080/2009-4 and 312394/2014-4, and FAPERJ grant 111.488/2013. RVM thanks grants: CNPq-306885/2013, Capes/Cofecub-2506/2015, Faperj/PAPDRJ-45/2013. J.I.B. Camargo acknowledges CNPq for a PQ2 fellowship (process number 308489/2013-6). We also acknowledge Steve Preston for the predictions of the occultations and the referee Lawrence H. Wasserman (Lowell Observatory) for his contributions to improve the text.

\label{lastpage}

\end{document}